\documentclass[12pt]{article}
\pdfoutput=1
\usepackage[utf8]{inputenc}
\usepackage{amsmath,amssymb,graphicx}
\usepackage{epsfig,color,float,cancel,cite}
\usepackage[titletoc,title]{appendix} 
\usepackage[export]{adjustbox}
\usepackage{multicol,tabu,multirow,bm}
\usepackage{hyperref}
\usepackage{slashed} 
\numberwithin{equation}{section}
\newcommand{\Kahler}{\ensuremath{\text{K}\ddot{\text{a}}\text{hler}\,}}
\def\bea{\begin{eqnarray}}
\def\eea{\end{eqnarray}}
\newcommand{\beq}{\begin{equation}}
\newcommand{\eeq}{\end{equation}}

\def\bal#1\eal{\begin{align}#1\end{align}}
\def\MeV{{\, \rm MeV}}

\def\TeV{{\, \rm TeV}}

\def\bel#1{\begin{equation} \label{#1}}
\def\ee{\end{equation}}
\allowdisplaybreaks
\begin{document} 
\thispagestyle{empty}
%
%
\begin{center}
\vspace*{0.5cm}
{\LARGE\bf Cosmology in the presence of multiple light moduli \\}
\bigskip
{\large  Bobby S. Acharya}\,$^{a,b,1}$, \, \, 
{\large  Mansi Dhuria}\,$^{c,2}$, \, \, 
{\large  Diptimoy Ghosh}\,$^{d,3}$, \, \, 
{\large  Anshuman Maharana}\,$^{e,4}$, \, \, 
{\large  Francesco Muia}\,$^{b,5}$ \, \,
\\
\bigskip 
\bigskip
{\small
$^a$Department of Physics, King's College London, London WC2R 2LS, UK \\[2mm]
$^b$International Centre for Theoretical Physics, Trieste 34014, Italy \\[2mm]
$^c$Institute of Infrastructure Technology Research and Management, Ahmedabad 380026, India \\[2mm]
\hspace*{-5mm} $^d$\mbox{Department of Physics, Indian Institute of Science Education and Research Pune, Pune 411008, India}\\[2mm]
$^e$Harish Chandra Research Institute, HBNI, Allahabad 211019, India
}
\end{center}
\bigskip 
\vspace*{1cm}
\begin{center} 
{\Large\bf Abstract} 
\end{center}
\vspace*{-0.35in}
\begin{quotation}
\noindent 
The generic expectation in string/supergravity models is that there are multiple moduli fields with masses of the order of the supersymmetry
breaking scale.
We study the cosmology that arises as a result of vacuum misalignment of these moduli fields (in contrast to previous studies
which mostly focussed on the single modulus case). We show that the dark radiation produced
from the heavier moduli undergoes significant dilution. This happens even if there is a small splitting between the masses of the 
lightest and the heavier moduli. On the other hand, in the absence of fast annihilation processes decay of heavier moduli generically
leads to overproduction of dark matter. We discuss a scenario where the problem can be addressed with a prompt dark matter annihilation
to dark radiation. This can lead to realistic dark matter abundances, and the additional dark radiation produced as a result of
this mechanism undergoes sufficient dilution as long as the annihilation is prompt.

\end{quotation}

\newpage 
\section{Introduction}

Moduli fields and hidden sectors are ubiquitous in string compactifications. It is important to develop an understanding of their implications
for phenomenology and examine the generic predictions. One such implication is {\it moduli cosmology} \cite{mod1, mod2, modr}.
For a large class of models, moduli fields experience vacuum misalignment as a result of inflation.  This leads to epoch(s) in the history of the
Universe where the energy density of the Universe is dominated by oscillating moduli fields. The Universe reheats with the decay of each species
of modulus field. In cases where one of the moduli
fields is significantly lighter than all others, it can be argued that the decay products of all the heavier moduli are diluted and the decay of the last modulus
effectively sets the initial conditions for the evolution of the Universe.  Most studies of moduli cosmology in the literature have focussed on this case.
However, in general one expects {\it multiple moduli} with masses close to the lightest one. This paper aims  to initiate the study of
moduli cosmology in such settings.
Our focus will be on dark radiation and dark matter abundances. The abundances will be sensitive to the branching ratios of the moduli to all
their decay products (both in the visible and hidden ones). We will quantify the size of branching ratios and the splittings necessary between
the light modulus and the heavier ones for viable phenomenology. 

In string/supergravity models the potential experienced by moduli fields depends on the inflaton vacuum expectation value (VEV) which results in
vacuum misalignment \cite{mismatch}. For a single modulus, $\varphi$, the dynamics in an  FRW background (after canonical normalisation) is
governed by the equation %
$$
\ddot{\varphi} + 3H \dot{\varphi} + {\partial V \over \partial{\varphi}} = 0 \,,
$$
where $H$ is the Hubble parameter and $V$ is the scalar potential. The form of the above equation implies
that the modulus remains pinned to its initial displacement $\varphi^{\rm in}$ till the Hubble constant falls below its mass. After this, it begins
to oscillate about its post-inflationary minimum; the associated energy density redshifts like matter. Finally the modulus decays, the Universe is injected
with its decay products. Typically moduli interact with Planck suppressed interactions and their lifetime is approximately given by 
$$
   t  \sim   { m^2_{\rm pl}  \over m_{\varphi}^3 } \, ,
$$
where $m_{\rm pl} = 2.43 \times 10^{18} \, \text{GeV}$ is the reduced Planck mass. Thus the amount of time available for the dilution of the decay products of a modulus is determined  by its mass. This makes the cosmology sensitive
to the moduli mass spectrum. To proceed, we assume that the mass spectrum is that of a generic model. Let us outline these assumptions (and the basis for these).
Moduli masses are determined by the potential
$$
     V = e^{{\mathcal{K}}} \left( \mathcal{K}^{i \bar{j}} D_{i} W D_{\bar{j}}  \overline{W} -  3 |W|^2 \right),
$$
where $\mathcal{K}$ and $\mathcal{W}$ are the \Kahler and superpotentials of the model and the derivatives are the \Kahler covariant derivatives.
The gravitino mass is given by 
$$
  m_{3/2} = \langle e^{\mathcal{K}/2} |W| \rangle \, .
$$
In the presence of a large number of hidden sectors\footnote{Consistency requirements
often force a large number of hidden sectors in string compactifications, see e.g.,  \cite{hidden, hreview}.},  one expects that supersymmetry breaking
takes place in a sector that communicates with the standard model via interactions of gravitational strength (see e.g \cite{MSUGRA, rewsbtwo, pk, douglas}).
In general there will also be moduli which interact with the supersymmetry breaking sector with interactions of the same strength; thus 
the generic expectation for these moduli is that their masses are approximately at the scale of the gravitino mass (which, in our case, is also the
mass scale of the visible sector superpartners). In summary, our setup  will consist of multiple moduli
with masses of order the lightest one\footnote{Situations where the lightest modulus is much lighter than all others can naturally arise if a
symmetry principle is operational such as in the Large Volume Scenario \cite{bala}.} with couplings of gravitational strength  to the visible and hidden sectors.

Dark radiation poses a serious challenge for moduli cosmology. Moduli decaying to radiation in hidden sectors or light axions lead to dark radiation.
Given the strong observational constraints on dark radiation \cite{planck}, these can be disastrous for phenomenology. Even in the case of one light
modulus, the constraints require that the constructions are special so that the dangerous decay channels are suppressed \cite{drst}.
The problem is amplified with multiple moduli, as each one of them can in principle decay to dark radiation. Tuning the constructions so that all
the dangerous channels have small branching ratios seems highly unnatural. In this work, we do not assume such unnatural tuning, and quantify
the splitting necessary between the lightest modulus and a heavy modulus so that dark radiation produced as a result of the decay of a heavier
modulus is sufficiently diluted. We find that the splitting necessary is not large, so need not be considered as imposing a strong constraint on
 the spectrum.

We next turn to dark matter. There are
various scenarios to produce dark matter in moduli cosmology (see e.g. \cite{mdm, ann, pk, modr}).  We will focus on the simplest case --  
dark matter produced as a result of decay of the moduli. We find that in the absence of annihilation processes, dark matter is overproduced
(the dilution mechanism for dark radiation discussed in the previous paragraph does not work for dark matter). We propose a viable scenario where
dark matter annihilates into dark radiation promptly
which in turn gets diluted by the mechanism discussed previously leading to realistic abundance.

\section{Two Moduli}

The essential features of our analysis can be captured by considering a model with two moduli. Thus, we carry out our analysis in the main
text considering a system with only two moduli (for completeness we discuss aspects of the many moduli case in Appendix \ref{Nmoduli}).
We denote the moduli by $\varphi_{\rm 1}$ and $\varphi_{\rm 2}$ and their masses by $m_1$ and $m_2$ respectively (with  $m_1 < m_2$).
Assuming only gravitational couplings, the life time of the moduli fields is of the order
$m_{\rm pl}^2/m_{\varphi_i}^3 \approx (16 \TeV / m_{\varphi_i})^3 \, \text{sec}$. This must be less than the age of the Universe at the time
of Big Bang Nucleosynthesis (BBN) so as not to spoil the successes of BBN \cite{mod1}. Thus $m_{1,2}$ must be
larger than $\sim$ 50 -100 TeV. We take $m_{1} \sim 50 \TeV$ as a benchmark value, and assume that $m_2$  is of the order of $m_1$ (but larger).

Following the arguments in \cite{mismatch}, we assume that during inflation the (canonically normalised) fields 
$\varphi_{1,\, 2}$ get a displacement of the order of $m_{\rm pl}$ from their late-time minima. After the end of inflation the Hubble parameter ($H$)
decreases with time. When $H \lesssim m_2 (m_1)$, $\varphi_2 (\varphi_1)$ starts oscillating around its minimum. 
We define the following quantities:
\begin{itemize}
\item[]  $\rho_{\rm 2,in} (\rho_{\rm 1,in})$ : energy density stored in the field $\varphi_2 (\varphi_1)$ when it starts oscillating;
\item[]  $H_{\rm 2,in} (H_{\rm 1,in})$ : Hubble parameter when $\varphi_2 (\varphi_1)$ starts oscillating; recall that $H_{\rm i,in} \sim m_i$;
\item[]  $a_{\rm 2,in} (a_{\rm 1,in})$ : scale factor when $\varphi_2 (\varphi_1)$ starts oscillating.
\end{itemize}
\begin{figure}[!h!]
\centering
\includegraphics[scale=0.5]{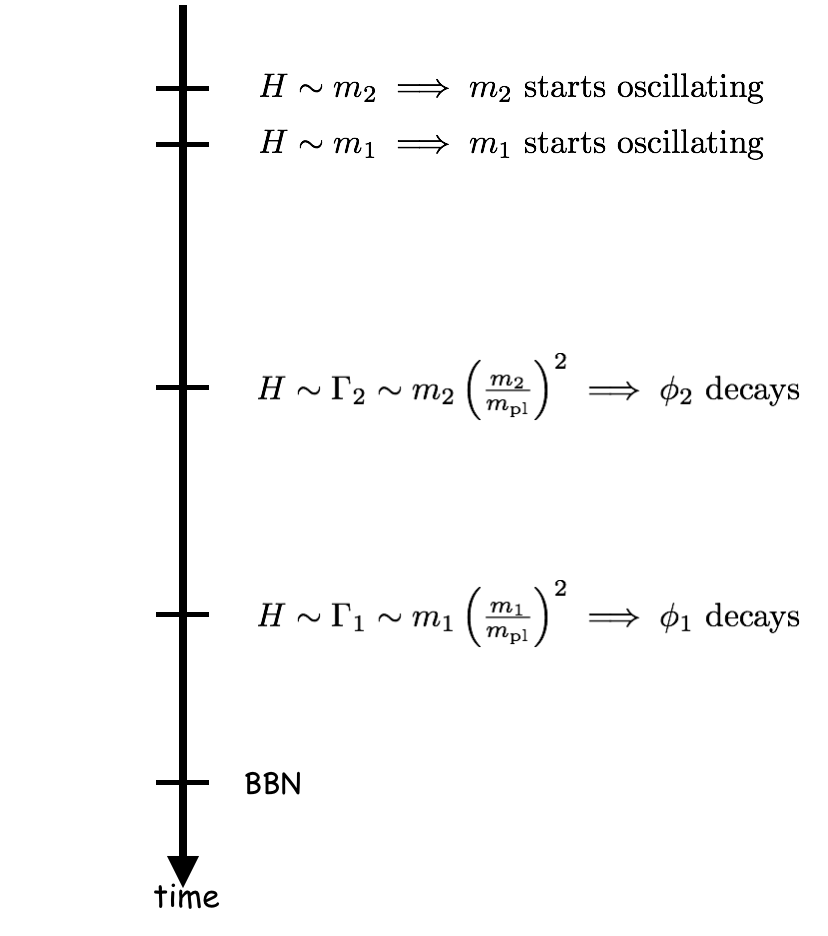}
\caption{A sketch of the time evolution of the Universe in the presence of two moduli.}
\end{figure}
Once $\varphi_2$ starts oscillating, the energy density associated with the initial displacement $\rho_{\rm 2, in}$ redshifts as matter, and quickly
dominates the energy density of the Universe. Thus the evolution is well approximated by assuming matter domination. 
When the Hubble parameter decreases to $H \sim \Gamma_2 \sim m_2^3/m_{\rm pl}^2$, the heavy modulus decays. Let us denote the
branching ratios of the decay of $\varphi_2$ to
dark matter, dark radiation, and the visible sector by ${\cal B}_{\rm 2,dm}$, ${\cal B}_{\rm 2,dr}$ and ${\cal B}_{\rm 2,vis}$ respectively. 
 The dark radiation and dark matter densities just after the decay of $\varphi_2$ are then given by\footnote{We assume the decay to be instantaneous.}
 
 \bea
\label{dr22}
\rho_{\rm dr}|_ {\varphi_2 \, {\rm decay}} &=& \rho_{2, {\rm in} } \left( \frac{\Gamma_2}{m_2} \right)^2  {\cal B}_{\rm 2,dr} \,, \\ 
\label{dm22-old}
\rho_{\rm dm}|_ {\varphi_2 \, {\rm decay}} &=& \rho_{2, {\rm in} } \left( \frac{\Gamma_2}{m_2} \right)^2  {\cal B}_{\rm 2,dm} \,, \\
\label{vis22}
\rho_{\rm vis}|_ {\varphi_2 \, {\rm decay}} &=& \rho_{2, {\rm in} } \left( \frac{\Gamma_2}{m_2} \right)^2  {\cal B}_{\rm 2,vis} \,,
\eea
where we have used the fact that, in the matter dominated era, the ratio of the scale factors at time $t_{A}$ and $t_{B}$ is related to the ratio
of the Hubble parameters  by
\bea
\left( \frac{a(t_B)}{a(t_A)} \right)^3 = \left( \frac{H(t_A)}{H(t_B)} \right)^2 \, .  
\eea

Note that, unless the mass of the dark matter particles are too close to $m_2/2$, they will be relativistic at the time of production.
In the set-up under consideration, it is natural to have the dark matter to arise from the hidden sector with a mass not much less than the
moduli mass scale (see for example, \cite{bob,bob1} for motivations for dark matter as hidden sector LSP). When needed, we will use a benchmark value
of approximately a TeV.

The number density of the dark matter particles at production can now be written as (assuming 2-body decay)
\bea
\label{dm22}
n_{\rm dm}|_ {\varphi_2 \, {\rm decay}} &=& \frac{2 \rho_{\rm dm}|_ {\varphi_2 \, {\rm decay}}}{m_2} \,.
\eea

Let us determine the equation of state of the Universe right after the decay of $\varphi_2$ ($H \sim \Gamma_2$).
At this stage,  the density of non-relativistic matter is  given by the contribution from the oscillations of the modulus $\varphi_1$ only.
This can be written as
\bea
\rho^{\rm mat}|_{\varphi_2 \rm{decay}}   &=& \rho_{\rm \textrm 1,  in} \left( \frac{\Gamma_2}{m_1} \right)^2 = \cr
&=& \frac{1}{2}m_{\rm pl}^2 m_1^2 \left( \frac{\Gamma_2}{m_1} \right)^2 = \cr
&=& \frac{1}{2}m_{\rm pl}^2 \Gamma_2^2  \,,
\label{matden}
\eea
where in going from the first line to the second, we have set the initial displacement of $\varphi_1$ to be of the order of $m_{\rm pl}$.
Similarly, the energy density of radiation at that time is 
\bea
\rho^{\rm rad}|_{\rm{ \varphi_2 decay}} &=& \rho_{\rm 2, in } \left( \frac{\Gamma_2}{m_2} \right)^2 \,
({\cal B}_{\rm 2,vis} + {\cal B}_{\rm 2,dr} + {\cal B}_{\rm 2,dm}) = \cr
&=& \frac{1}{2}m_{\rm pl}^2 m_2^2 \left( \frac{\Gamma_2}{m_2} \right)^2 \,
({\cal B}_{\rm 2,vis} + {\cal B}_{\rm 2,dr} + {\cal B}_{\rm 2,dm}) = \cr
&=&    \frac{1}{2}m_{\rm pl}^2  \Gamma_2^2 \, ({\cal B}_{\rm 2,vis} + {\cal B}_{\rm 2,dr} + {\cal B}_{\rm 2,dm}) \,.
\label{radden} 
\eea
Comparing Eq. \eqref{matden} and Eq. \eqref{radden}, we see that  the Universe is at an epoch of matter-radiation equality
at $\varphi_2$ decay, and thus becomes matter dominated soon after the decay of $\varphi_2$.

After the decay of $\varphi_2$, the Universe evolves adiabatically until the decay of $\varphi_1$. This occurs when $H \sim \Gamma_1$.
Given the results of the previous paragraph, we take the epoch between the decay of the two moduli to be matter dominated.
As described in the introduction, a sizeable branching fraction of the decay of $\varphi_1$ to dark radiation is disastrous, thus we
assume that this is negligible. However, the contribution from the decay of the heavier moduli can in principle also be significant, and demanding negligible
branching ratios of all the heavier moduli to dark radiation seems too stringent a requirement. Thus it is natural to take a non-negligible branching ratio
of the heavier modulus to dark radiation, and study possible mechanisms 
so as to get viable phenomenology.

Let us denote the branching ratios of $\varphi_1$ to dark matter and visible sectors by ${\cal B}_{\rm 1,dm}$ and ${\cal B}_{\rm 1,vis}$ respectively.
We will now compute the energy densities of the various components after the decay of $\varphi_1$. As these are the energy densities
right after the last reheating epoch, we denote them with a superscript ${\textrm{{``rh"}}}$.

 \noindent \underline{Dark radiation:} Dark radiation dilutes as $a^{-4}(t)$ in the epoch between the decay of the two moduli. Making use of Eq. \eqref{dr22}
we have (see appendix~\ref{app-A})

\bel{drrh}
\rho_{\rm dr}^{\rm rh} = \rho_{\rm dr}|_ {\rm \varphi_2 \, decay} \left( \frac{\Gamma_1}{\Gamma_2} \right)^{8/3}  = 
\rho_{\rm 2, in} \left( \frac{\Gamma_2}{m_2} \right)^2  {\cal B}_{\rm 2,dr}    \left( \frac{\Gamma_1}{\Gamma_2} \right)^{8/3} \,. \\ 
\ee

\noindent \underline{Dark Matter:} Dark matter receives contributions from the decay of both the moduli. Since, there is no annihilation the
number density of dark matter produced from the decay of $\varphi_2$ dilutes as $a^{-3}(t)$ in the epoch between the decay of the two moduli.
Thus, using Eq. \eqref{dm22} we find that the contribution to the dark matter number density from the decay of $\varphi_2$ at the time of $\varphi_1$ decay is given by 
\bea
n_{\rm dm}^ {(2), \rm rh} &=&  n_{\rm dm}|_ {\rm \varphi_2 \, decay}  \left( \frac{\Gamma_1}{\Gamma_2} \right)^2 = \cr
&=& \rho_{\rm 2, in} \left( \frac{\Gamma_2}{m_2} \right)^2  {\cal B}_{\rm 2,dm}  \left( \frac{\Gamma_1}{\Gamma_2} \right)^2 \frac{2}{m_2} = \cr
&=&  \frac{1}{2} m_{\rm pl}^2 \, \Gamma_1^2  \, {\cal B}_{\rm 2,dm} \frac{2}{m_2} \,. \label{nDM2rh}
\eea

Similarly, the contribution from the decay of $\varphi_1$ is
\bea
\label{dm-justify}
n_{\rm dm}^ {(1), \rm rh} &=&  \frac{\rho_{\rm 1, in}  \left( \frac{\Gamma_1}{m_1} \right)^2 \, {\cal B}_{\rm 1,dm} }{m_1/2} = \cr
&=&  \frac{1}{2} m_{\rm pl}^2 \, \Gamma_1^2  \, {\cal B}_{\rm 1,dm} \frac{2}{m_1} \,.
\eea
Note that there is a mild hierarchy, $m_1/m_2$, between the two contributions. 
Now, let us turn to the visible sector.

\noindent \underline{Visible Sector:} The visible sector gets contributions from the decay of both the moduli. It dilutes as $a^{-4}(t)$ in the epoch between the 
decays of the two moduli. Using Eq. \eqref{vis22}, we find
 \bea
\rho_{\rm vis}^{\rm rh} &=&  \rho_{\rm vis}|_ {\rm \varphi_2 \, decay} \left( \frac{\Gamma_1}{\Gamma_2} \right)^{8/3} +
\rho_{\rm 1,osc}|_ {\rm \varphi_1 \, decay}  \, {\cal B}_{\rm 1,vis} = \\
&=& \rho_{\rm 2, in} \left( \frac{\Gamma_2}{m_2} \right)^2  {\cal B}_{\rm 2,vis}    \left( \frac{\Gamma_1}{\Gamma_2} \right)^{8/3} +
\rho_{\rm 1, in}  \left( \frac{\Gamma_1}{m_1} \right)^2  {\cal B}_{\rm 1,vis}  \,.
 \label{visrh}
\eea
Taking again the initial displacements of the fields to be of the order $m_{\rm pl}$,  the reheating temperature of the visible sector can be
easily obtained from Eq. \eqref{visrh}. We get
\bea
\label{vt}
T_{\rm rh} =  \left( \frac{\rho_{\rm vis}^{\rm rh} }{\hat{g}_{\rm vis}(T_{\rm rh} ) \, \pi^2/30} \right)^{1/4} \approx 13 \MeV 
\left( \frac{{\cal B}_{\rm 1,vis}+ (m_1/m_2)^2 {\cal B}_{\rm 2,vis}}{\hat{g}_{\rm vis}(T_{\rm rh})/10}  \right)^{1/4} \left(\frac{m_1}{100 \TeV}\right)^{3/2} \,.
\eea
Note that, the contribution from the decay of $\varphi_2$ is smaller than that from the decay of $\varphi_1$ by a factor of
$\left( \frac{\Gamma_1}{\Gamma_2} \right)^{2/3} \sim \left ({ m_1 \over m_2} \right)^2$. 

\section{Observables}

In the previous section, we have computed three energy densities: the energy density of the visible sector at the end of reheating $(\rho^{\rm rh}_{\rm vis})$,
the energy density of dark radiation at the end of reheating $(\rho^{\rm rh}_{\rm dr})$ and the energy density of dark matter at the end of reheating
$(\rho^{\rm rh}_{\rm dm})$. Next, let us use these to compute cosmological observables in the `two moduli' model.

\subsection{$\Delta N_{\rm eff}$ at the time of decoupling of the neutrinos} 
\label{droo}

The quantity $\Delta N_{\rm eff}$ at the time of decoupling of the neutrinos is defined to be the ratio of the energy density in dark radiation and
the energy density in one species of neutrino at the time of decoupling,
\bel{ndef}
  \Delta N_{\rm eff}  = { 3  \rho_{\rm dr} ( t_{\nu}) \over \rho_{\nu}  (t_{\nu}) } =  { 43  \over 7 } {   \rho_{\rm dr} ( t_{\nu}) \over \rho_{\rm vis}  (t_{\nu}) },
\ee
where $t_{\nu}$ denotes the time of decoupling of neutrinos. Let us begin by expressing this in terms of  $\rho_{\rm rh}^{\rm vis}$ and $\rho_{\rm rh}^{\rm dr}$
(which we have computed in the previous section). Assuming no  significant change in the number of degrees of freedom in dark radiation between
the time of last reheating and the decoupling of neutrinos, we have:
\bel{abc}
  \rho_{\rm dr}(t_{\nu}) =   \rho_{\rm dr}^{\rm rh} { a^{4} (t_{\rm rh} ) \over a^{4} ( t_{\rm \nu} ) }.
\ee
We also have for the visible sector
\bel{visr}
   \rho_{\rm vis}(t_{\nu}) = \rho_{\rm vis}^{\rm rh} {  \hat{g} ( T_{\nu} )  T_{\nu}^4 \over \hat{g}( T_{\rm rh}) T_{\rm rh}^4 } \,,
\ee
where $T_{\nu}$ is the temperature of the visible sector\footnote{All temperatures and the $\hat g$ factor quoted are for the visible sector.} at the time of
decoupling of the neutrinos, and $\hat g (T)$ is the effective number of relativistic degrees of freedom in the visible sector at temperature $T$.
Ratios of the temperatures can be related to the ratio of the scale factors by making use of the conservation of
entropy in the visible sector,
\bel{ento}
a^{3}(t_{\rm rh}) \hat g (T_{\rm rh}) T^3_{\rm rh}  = a^{3}(t_{\rm \nu}) \hat g (T_{\rm \nu}) T^3_{\rm \nu} \, .
\ee
Combining Eq.s \eqref{ndef}, \eqref{abc}, \eqref{visr}, and \eqref{ento} we have
\bel{drint}
 \Delta N_{\rm eff} = { 43 \over 7}  {  \rho_{\rm dr}^{\rm rh} \over  \rho_{\rm vis}^{\rm rh} } { {\hat g}^{1/3} ( T_{\nu} ) \over
 {\hat g}^{1/3} (T_{\rm rh} ) } \, .
 \ee
%
Plugging in the results from the previous section and taking $\hat g(T_{\rm rh}) \approx \hat g(T_\nu) \approx 10$, one finds (taking the leading
contribution i.e., the second term in Eq.~\ref{visrh})
\bel{numneff}
   \Delta N_{\rm eff} = { 43 \over 7}  {{ \rho_{\rm 2,in}  \left( \frac{\Gamma_2}{m_2} \right)^2  {\cal B}_{\rm 2,dr}    \left( \frac{\Gamma_1}{\Gamma_2} \right)^{8/3}} \over {\rho_{\rm 1,in}  \left( \frac{\Gamma_1}{m_1} \right)^2  {\cal B}_{1,vis}} }.
 \ee
Taking the displacements to be of the order of $m_{\rm pl}$, we get 
\bel{numneffin}
  \Delta N_{\rm eff}  \approx  6.14 {  {\cal B}_{\rm 2,dr}  \over {\cal B}_{1,vis} } \left( { m_1 \over m_2} \right)^2.
\ee
Alternatively,
\bel{ratfin} 
  m_2 =    {2.48 \over { \sqrt{\Delta N_{\rm eff}  } } }\sqrt{ {  {\cal B}_{\rm 2,dr}  \over {\cal B}_{\rm1,vis} } } m_1.
\ee
%
This implies that the dark radiation produced from the decay of heavier moduli can be diluted so as to be consistent with
the observations even if the mass of $\varphi_2$ is only a few times that of the mass of $\varphi_1$. 
For example, $m_2 \sim 5 m_1$ leads to $\Delta N_{\rm eff} \lesssim 0.25$ (for ${\mathcal O (1})$ branching ratios) which is
consistent with the current bound from CMB \cite{planck}. The mechanism is quite robust; even $\Delta N_{\rm eff} \lesssim 0.03$, 
which might be achieved by the CMB Stage-IV experiment \cite{Abazajian:2016yjj}, requires $m_2 \sim 15 m_1$.
A similar conclusion
holds for cases with multiple moduli, which we discuss in Appendix~\ref{Nmoduli}. Thus, a small split in the moduli spectrum can prevent
dark radiation overproduction.

\subsection{Dark matter abundance}

As described in the previous section, dark matter is produced from the decay of both the moduli. Let us examine the contribution
from the decay of the heavier modulus to the dark matter abundance if there is no annihilation. To do this, we compute
the ratio of number density of dark matter produced from the decay of the heavier modulus and the visible sector entropy density.
Using Eq. \eqref{nDM2rh}  and 
\begin{equation}
\label{sfor}
s = {2 \pi^{2} \over 45} g_{\rm vis} (T_{\rm rh}) T_{\rm rh}^{3} \,,
\end{equation}
we get, 
\begin{equation}
\label{ns}
 { n^{(2), \rm{rh}}_{\rm dm} \over s } = {3 \over 2} { \hat{g}_{\rm vis}(T_{\rm rh}) \over {g}_{\rm vis}(T_{\rm rh}) } {T_{\rm rh} \over m_{1} } \,
  \frac{{\cal B}_{\rm 2, dm }}{{\cal B}_{\rm 1, vis } }  \left( {m_1 \over
 m_2} \right) \,.
\end{equation}
Recall that observations set
\begin{equation}
\label{dmo}
  {n_{\rm dm} \over s }\bigg{|}_{\rm Obs} = 5 \times 10^{-12} \left( { {\rm 100~GeV}  \over  m_{\chi} } \right) \,,
\end{equation}
where $m_\chi$ is the mass of the dark matter. 
Taking $m_1 \approx 100 \, \rm TeV$, $m_\chi \approx 100 \, \rm GeV$ and the reheat temperature from Eq. \eqref{vt},  we find
that 
\begin{equation}
\label{ns2}
 { n^{(2), \rm{rh}}_{\rm dm} \over s } \approx  10^{5}  \, \frac{{\cal B}_{\rm 2, dm }}{{\cal B}_{\rm 1, vis } }  \left( {m_1 \over
 m_2} \right)    {n_{\rm dm} \over s }\bigg{|}_{\rm Obs}  \, .
 \end{equation}
For $m_1/m_2 \sim 0.1$, the contribution from the heavy modulus is approximately four orders of magnitude larger than the
observed value.

\section{Dark matter dilution mechanism}
\label{dmdilm}

As we have seen in the previous section, in the absence of annihilation processes  dark matter  is generically overproduced.
It is interesting to ask whether there is a natural way to reduce the abundance of dark matter produced form the decay of the heavier
moduli. 
We will argue that such a mechanism exists if the dark matter produced can annihilate \footnote{Annihilation of dark matter produced from decay of moduli
has been previously invoked in \cite{ann, pk}.} to dark radiation. This is facilitated if the hidden sector from which the dark matter arises also contains dark radiation. 
The dark matter  that survives after the self-annihilation can have the correct abundance. Additional dark radiation is produced, but this is diluted 
as described before, if the annihilation processes are prompt and cease well before the decay of the lightest modulus.

 Since dark matter annihilation has to be such that the dark radiation produced gets diluted, we will demand that
 $\langle\sigma v\rangle_{\rm{ann}}$ of the annihilation process is large enough so that the dark matter produced immediately
 annihilates into dark radiation. This is the case if 
\begin{equation}
\label{condii}
n_{\rm dm}|_ {\varphi_2 \, {\rm decay}} \langle\sigma v \rangle_{\rm{ann}} > H(t_{\varphi_{\rm 2} \rm{decay}}) \,.
\end{equation}

Making use of Eq. \eqref{dm22} in Eq. \eqref{condii} we find 
\begin{equation}
 \mathcal{B}_{2, \rm dm} m_{2}^2\langle \sigma v \rangle_{\rm{ann}} > 1 \,.
\end{equation}
The annihilation stops approximately when the annihilation rate becomes of order of Hubble. At that point, the number density, ${\hat n}_{\rm dm}$, is given
by  
\begin{equation}
{\hat n}_{\rm dm}  \simeq \frac{H(t_{\varphi_{\rm 2} \rm{decay}})}{\langle\sigma v \rangle_{\rm{ann}}} \simeq
\frac{ m_{2}^3}{\langle \sigma v \rangle_{\rm ann} \, m_{\rm pl}^2} \, .
\end{equation}
Redshifting  this abundance till the decay of the lightest modulus gives
\begin{equation}
{\hat n}_{\rm dm}^{(2), \rm rh}  \simeq \frac{ m_{1}^3}{\langle \sigma v \rangle_{\rm ann} \, m_{\rm pl}^2} \left(\frac{m_1}{m_2}\right)^3 \,.
\end{equation}
 This has to be compared to
the contribution of the decay of the heavy modulus to dark matter number density at the time of decay of the light one without the annihilation
process (which we defined
to be  $n_{\rm dm}^{\rm(2)  rh} $ in Eq. \eqref{nDM2rh}),
\begin{equation}
n_{\rm dm}^{\rm(2), rh}  \simeq \mathcal{B}_{2, \rm dm} { {m_1}^6 \over {m_{\rm pl}^2} } {1 \over m_2}\,.
\end{equation}
The ratio is 
\begin{equation}
\label{eq:RatioAbundances}
\frac{{\hat n}_{\rm dm}^{(2), \rm rh}}{n_{\rm dm}^{\rm(2), rh} } 
 \simeq
\frac{1}{\mathcal{B}_{2, \rm dm}} \, \frac{1}{m_2^2 \langle \sigma v \rangle_{\rm ann}}  \, .
\end{equation}
Dark matter produced is expected to become non-relativistic very quickly \cite{ann,Arcadi:2011ev}. In this case, if one roughly estimates
the cross section to be $\langle \sigma v \rangle_{\rm ann} \simeq g^4/m_{\chi}^2$ then
\begin{equation}
\label{eq:RatioAbundances2}
\frac{{\hat n}_{\rm dm}^{(2), \rm rh}}{n_{\rm dm}^{\rm(2), rh} } \simeq 
 \frac{1}{g^4 \,\mathcal{B}_{2, \rm dm}} \, \left(\frac{m_\chi}{m_{2}}\right)^2  \, .
\end{equation}
Thus, the dark matter overproduction problem can be alleviated if there is approximately two orders of hierarchy between
$m_\chi$ and $m_{2}$ (which can be satisfied if $m_\chi \sim$ 100 GeV - TeV as taken in the previous section). Such a hierarchy
can be obtained if there is a loop suppression between the masses generated for the fermions and the scalars, see for example
\cite{Randall:1998uk,Acharya:2008zi,bob}.
Since the ratio in Eq. \eqref{eq:RatioAbundances} scales inversely with the annihilation cross-section, having dark
radiation in the same sector as the dark matter helps reducing the dark matter abundance.

\section{Conclusions}

Moduli fields are a generic feature of string compactifications.  Their vacuum misalignment 
can have a significant impact on cosmology. Another generic feature of string compactifications is the existence of
hidden sectors. It is important to understand if these two ingredients have any implications
for cosmology. As it is natural to expect multiple moduli at the gravitino mass scale, we have
initiated the study of modular cosmology in the presence of multiple moduli. We have focused  on
dark radiation and dark matter abundances. For dark radiation, even a small split between the lightest
modulus and the heavier ones dilutes the radiation produced from the heavier moduli, so that
they do not lead to overproduction of dark radiation. For dark matter, contributions from the 
decay of heavier moduli leads to overproduction in the absence of annihilation processes. We 
have seen that there is an natural solution to this problem -- fast annihilation of the dark matter to
dark radiation in its sector. With this, if the heavier moduli with sizable branching fractions to dark
matter are two or three orders of magnitude heavier than the dark matter particle, then there is no over
production of dark matter. This mild hierarchy seems to be necessary for viable phenomenology.

Of course, there are many  directions to explore. 
It will be interesting to study in detail the implications of our scenario for direct detection experiments.  Our analysis has
relied on generic expectations on the spectrum and initial displacements of the moduli fields, and  it would be interesting to carry out
studies of explicit string constructions. 

\subsection*{Acknowledgements}

We would like to thank Fernando Quevedo for useful discussions. 
MD, DG and AM acknowledge hospitality of the HECAP section of ICTP Trieste where large part of this work was done.
BA was supported by a grant  ($\#$ 488569, Bobby Acharya) from the Simons Foundation.
MD would like to acknowledge support through Inspire Faculty Fellowship of the Department of Science and Technology, Government
of India under the Grant Agreement number: IFA18-PH215. DG and AM would also like to acknowledge supports through
Ramanujan Fellowships of the Department of Science and Technology, Government of India.  

\appendix

\section{Multiple Moduli}
\label{Nmoduli}
It is easy to generalise our discussion to $N>2$ moduli (with N not very large). For $N$ moduli $\varphi_i$ of mass $m_i$, the entire epoch
between the oscillation of the first modulus to the decay of the last modulus can be approximated to be matter dominated. If each of the
modulus has order one branching fraction to the visible sector, then the visible sector energy density at the time of the last reheating is dominated
by the contribution from the decay of the lightest modulus. 

Dark radiation produced from the decay of the $i^{\rm th}$ modulus make a contribution to $\Delta N_{eff}$ (at the time of decoupling
of neutrinos) which is given by
$$
   \Delta N^{i}_{\rm eff}  \approx  6.14 {  {\cal B}_{\rm i,dr}  \over {\cal B}_{\rm 1,vis} } \left( { m_1 \over m_i} \right)^2,
$$
where ${\cal B}_{\rm i,dr}$ is the branching fraction of the $i^{\rm th}$ modulus to dark radiation. Thus, the contributions
from the heavier moduli drop inversely with square of their masses, and do not pose a serious problem for $N_{\rm eff}$.

In the absence of annihilation processes, dark matter produced from the decay of the $i^{\rm th}$ modulus contributes to
the dark matter number density at the time of the last reheating by an amount
\bea
n_{\rm dm}^ {(i), \rm rh} &=&  n_{\rm dm}|_ {\rm \varphi_i \, decay}  \left( \frac{\Gamma_1}{\Gamma_i} \right)^2 \cr
&=& \rho_{\rm i, in} \left( \frac{\Gamma_i}{m_i} \right)^2  {\cal B}_{\rm i,dm}  \left( \frac{\Gamma_1}{\Gamma_i} \right)^2 \frac{2}{m_i} \cr
&=&  \frac{1}{2} m_{\rm pl}^2 \, \Gamma_1^2  \, {\cal B}_{\rm i,dm} \frac{2}{m_i}  \label{nDMirh}
\eea
Evolving this to the time of decay of $\varphi_{1}$, the ratio of this contribution to the entropy density
is 
\begin{equation}
\label{nsii}
 { n^{(i), \rm{rh}}_{\rm dm} \over s } \approx  10^{5}  \, \frac{{\cal B}_{\rm i, dm }}{{\cal B}_{\rm 1, vis } }  \left( {m_1 \over
 m_i} \right)    {n_{\rm dm} \over s }\bigg{|}_{\rm Obs}  \, ,
 \end{equation}
which leads to dark matter over abundance. On the other hand, in the presence of annihilation processes, as described in section
\ref{dmdilm}, the number density at the time of the last reheating (${\hat n}_{\rm dm}^{(i), \rm rh}$) is reduced,  its ratio to ${ n^{(i), \rm{rh}}_{\rm dm} }$ is
\begin{equation}
\label{eq:RatioAbundancesappi}
\frac{{\hat n}_{\rm dm}^{(i), \rm rh}}{n_{\rm dm}^{\rm(i), rh} } 
 \simeq
\frac{1}{\mathcal{B}_{i, \rm dm}} \, \frac{1}{m_i^2 \langle \sigma v \rangle_{\rm ann}}  \, .
\end{equation}
Thus, in summary, the qualitative picture discussed in the two moduli case does not change in the presence of many moduli.
\section{Ratio of scale factors in terms of ratio of Hubble parameters}
\label{app-A}

%
%
\subsection{Matter dominated era}
In the matter dominated era, we have
\bea
a &\propto& t^{2/3} \\
\implies H &=& \frac{1}{a} \frac{da}{dt} = \frac{2}{3} \frac{1}{t} 
\eea

Thus, 
\bea
\implies \frac{da}{a} &=& \frac{2}{3} \frac{dt}{t}
\eea
which, after integration, gives
\bea
{\rm Log} (a_f/a_i) &=& \frac{2}{3} \, {\rm Log} (t_f/t_i) = \frac{2}{3} \, {\rm Log} (H_i/H_f) \\
\implies \left( \frac{a_f}{a_i} \right)^3 &=&  \left( \frac{H_i}{H_f} \right)^2 \, , \,  \left( \frac{a_f}{a_i} \right)^4 =  \left( \frac{H_i}{H_f} \right)^{8/3}
\eea

\subsection{Radiation dominated era}
In the radiation dominated era, we instead have
\bea
a &\propto& t^{1/2} \\
\implies \left( \frac{a_f}{a_i} \right)^3 &=&  \left( \frac{H_i}{H_f} \right)^{3/2} \, , \,  \left( \frac{a_f}{a_i} \right)^4 =  \left( \frac{H_i}{H_f} \right)^2
\eea

%

\end{document}